# Twisted Angle-Dependent Work Functions in CVD-Grown Twisted Bilayer Graphene by Kelvin Probe Force Microscopy

Shangzhi Gu[1,2], Wenyu Liu[1], Guoyu Xian[2], Shuo Mi[1], Jiangfeng Guo[1], Songyang Li[1], Le Lei[1], Haoyu Dong[1], Rui Xu[1], Fei Pang[1], Shanshan Chen[1], Haitao Yang[2], Zhihai Cheng[1, *]

[1] *Beijing Key Laboratory of Optoelectronic Functional Materials & Micro-nano Devices, Department of Physics, Renmin University of China, Beijing 100872, China.*
[2] *Beijing National Laboratory for Condensed Matter Physics, Institute of Physics, Chinese Academy of Sciences, P.O. Box 603, Beijing 100190, China.*

**Abstract:** Tailoring the interlayer twist angle of bilayer graphene (BLG) has a significant influence on its electronic properties, including superconductivity, topological transitions, ferromagnetic states and correlated insulating states. These exotic electronic properties are sensitively dependent on the work functions of bilayer graphene samples. Here, the twisted angle-dependent work functions of CVD-grown twisted bilayer graphene (tBLG) are detailed investigated by Kelvin Probe Force Microscopy (KPFM) in combination with Raman spectra. The thickness-dependent surface potentials of Bernal-stacked multilayer graphene were measured. The AB-BLG and tBLG are directly determined by KPFM due to their twist angle-specific surface potentials. The detailed relationship of twist angles and surface potentials are further obtained by the *in-situ* combination investigation of KPFM and Raman spectra measurements. The thermal stability of tBLG was further explored through controlled annealing process. Our work provides the twisted angle-dependent surface potentials of tBLG and lays the foundation for further exploring their twist-angle-dependent novel electronic properties.

* To whom correspondence should be addressed: zhihaicheng@ruc.edu.cn

**Introduction**

As a particular two-dimensional (2D) material, graphene has attracted much attention due to its unique electronic, optical and mechanical properties. The electronic properties of the graphene are well known to be strongly dependent on the stacking orientations of layers. For instance, a dramatic change of the electronic band structure is appearing in the twisted bilayer graphene (tBLG). The tBLG was proven to exhibit distinct physics that was different from Bernal (AB) stacking bilayer graphene (BLG), such as the Moiré patterns resulting from the twisted stacking layers at a small angle. In particular, near a small magic twist angle, bilayer graphene transforms from a weakly correlated Fermi liquid to a strongly correlated two-dimensional electron system with exotic properties of unconventional superconductivity [1-3], topological transitions [4], magnetic transport characteristics [5], Mott correlated electronic states [6, 7], etc.

Among several methods for preparing 2D materials, chemical vapor deposition (CVD) is considered as one of the most popular methods. The CVD-grown graphene films have been successfully prepared with specific thickness and high crystalline quality [8, 9]. Thus, the growth of tBLG by CVD technique would be important for the investigation of their angle-dependent physical properties and potential applications [10-15]. In addition, it's also critical to directly characterize their angle-dependent structural characteristics. Recently, transmission electron microscopy (TEM) [16, 17], micro-Raman spectroscopy [18, 19], and various scanning probe microscopy (SPM) techniques [20-26] have been widely used to study the Moiré patterns of the tBLG samples constructed by the artificial stacking methods.

For CVD-grown twisted graphene layers, the angle-dependent structural and vibrational characteristics of tBLG have been revealed by Raman spectra measurements [27-31]. From the spatial resolution perspective, the SPM-based approaches are rather promising in the characterization of their electronic properties. A nonmonotonic angle-dependent vertical electrical conductivity across the interface of tBLG have been discovered by conductive atomic force microscopy (C-AFM) [32]. The electronical properties of tBLG films are sensitively dependent on their local work functions or surface potentials. Then, as a starting point, there is an urgent need to determine the rotation-relating bilayer domains with micron-scale resolution and quantitatively measure their angle-dependent work functions in the CVD-grown tBLG films. As for the measurements of surface potentials or work functions, Kelvin Probe Force Microscopy (KPFM) has been proven as an excellent technique with outstanding resolution and sensitivity [33, 34].

In this paper, we present a comprehensive KPFM investigation of the surface potential properties of CVD-grown multilayer graphene films with varying thickness and twist angles in combination with Raman spectra. The CVD-grown graphene films were characterized on the SiO$_2$/Si substrate via a transfer process after the direct growth on copper foils. The thickness-dependent surface potentials were first determined for the Bernal-stacked multilayer graphene films. The AB-BLG and tBLG flakes or domain are directly determined by their different surface potentials. The twisted angle-dependent surface potentials of tBLG are further detailed investigated based on an *in-situ* combination study of KPFM and Raman spectra measurements. The thermal stability of tBLG was also discussed via controlled sequential annealing process.

**Results and Discussion**

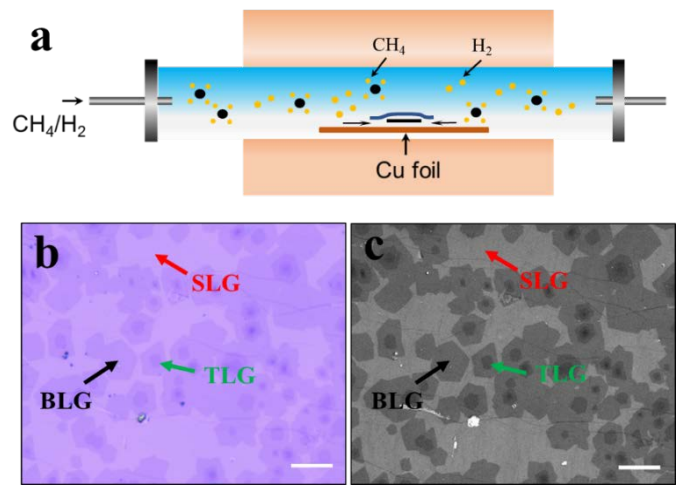

**Figure 1. Optical and SEM characterization of the CVD-grown multilayer graphene films.** (**a**) Schematic for the CVD growth process of multilayer graphene on the copper foil with the underlayer growth mode. The new graphene layer (blue) nucleates below the single layer (black). (**b**) Optical image of a multilayer CVD graphene films transferred on the SiO$_2$/Si substrate. (**c**) The corresponding SEM image of multilayer graphene in (b). The single layer (SLG), bilayer (BLG), trilayer (TLG) graphene regions are highlighted by the arrows and determined according to their relative contrast. No interior crystalline and twisted domains of the multilayer graphene films are visualized in (b) and (c). Scale bar is 70 µm in (b) and (c).

The multilayer graphene films were synthesised by a modified low-pressure CVD system [15]. A schematically illustration of the experimental CVD set-up is presented in Fig. 1(a). By controlling growth conditions, high-quality graphene grains with a size of several hundred microns were obtained. For multilayer graphene, the second layer would form beneath the first layer due to the reaction of catalysed copper surfaces. This underlying growth mode of multilayer graphene films is illustrated in Fig. 1(a). After growth, the large-area graphene films were transferred on the $SiO_2$/Si substrate using wet transfer process [35, 36].

Fig. 1(b) shows the optical image of one typical multilayer graphene films on the $SiO_2$/Si substrate. The single layer (SLG), bilayer (BLG), trilayer (TLG) graphene regions were clearly resolved based on their specific optical contrast. The top graphene layer is continuing and fully covered the whole $SiO_2$/Si substrate. For the partial BLG (and TLG) regions, the second layer (and third layer) is directly on the bottom substrate and under the top single layer (and bilayer). Fig. 1(c) shows the corresponding scanning electron microscopy (SEM) image of the multilayer graphene films in Fig. 1(b). It provides the layer number contrast and distributions between SLG, BLG, and TLG regions. Even with differential-interference-contrast optical images and SEM images, the interior crystalline and twisted domains of the multilayer graphene films are not discernible.

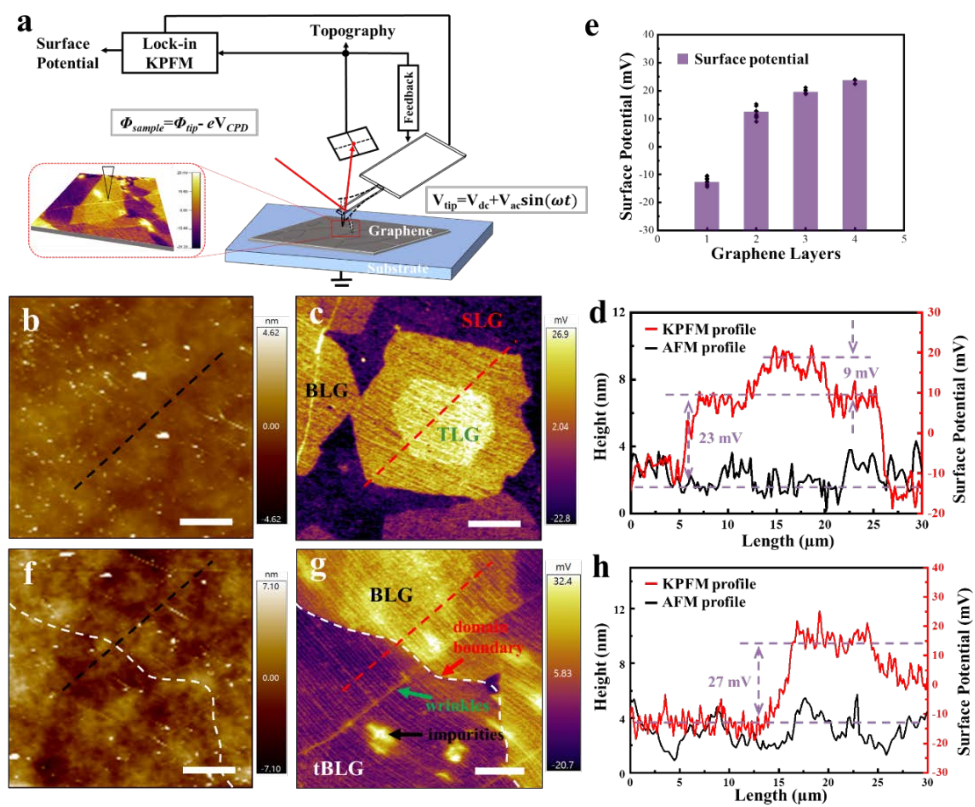

**Figure 2. KPFM characterization of the CVD-grown multilayer graphene transferred on $SiO_2$/Si substrate.** (**a**) Schematic illustration of the KPFM setup. (**b**) Topography image of a conventional multilayer graphene region. (**c**) The corresponding surface potential image of (b) taken by KPFM. (**d**) Cross-section profiles of topography in (a) and surface potential in (c). (**e**) Histogram of the averaged surface potentials of Bernal-stacked monolayer, bilayer, trilayer and tetralayer graphene. (**f**) Topography

image of a bilayer graphene region. (**g**) The corresponding surface potential image of (f). Two specific areas with different surface potentials are only clearly resolved in (g), which are determined as AB-BLG and tBLG, respectively. The interior grain boundary of the bilayer graphene region is highlighted by the white dashed lines in (f) and (g). The green and black arrows indicate the typical graphene wrinkles and surface contaminations. (**h**) Cross-section profiles of topography in (f) and surface potential in (g). Scale bar is 6 µm in all AFM images.

Many electrical SPM techniques, such as scanning microwave impedance microscopy (sMIM), electrostatic force microscopy (EFM), C-AFM and KPFM, have been used to investigate the local electrical properties of two-dimensional materials [32, 37-39]. KPFM is an effective AFM technique, which can spatially map and study surface topography and local work function variations simultaneously in ambient conditions [40-42]. At here, the electrical characterizations of CVD-grown multilayer graphene films were further performed by KPFM to measure their local work functions. Work function is an intrinsic band property of the material, which quantifies the minimum energy required to bring an electron from the Fermi level ($E_F$) to the vacuum level ($E_V$).

As shown by the schematic illustration of KPFM in Fig. 2(a), a conductive AFM tip is electrically biased (DC+AC) against a grounded sample. The bias-dependent electrostatic force between the tip and the sample was measured and minimized by adjusting the DC via the KPFM feedback. The measured DC voltage of the bias tip, corresponding to contact potential differences (CPD), determines the work functions difference between the AFM tip and the targeted region of the graphene films. The CPD is linked the work function (WF) of tip and sample with the following relationship, $V_{CPD}$= (WF$_{sample}$ - WF$_{tip}$)/$e$, where $e$ is elementary charge unit. The opposite value of CPD is determined as surface potential, which is used in the following.

Fig. 2(b) shows the AFM topography of a typical multilayer graphene region, including SLG, BLG and TLG, selected via optical imaging. The multilayer areas can be confirmed AB-stacked configurations by Raman features. It is noted that the thickness of graphene layers is not distinguishable in the topography image, which is due to the underlying growth mode and the surface roughness of the SiO$_2$/Si substrate. While the SLG, BLG and TLG regions are clearly resolved by their specific surface potentials, as shown in Fig. 2(c). The corresponding line profiles of the topography and surface potential, as indicated in Fig. 2(b) and (c), are shown in Fig. 2(d). The surface potential of TLG (and BLG) is ~9 mV (~23 mV) higher than that of BLG (and SLG). Fig. 2(e) shows the thickness-dependent surface potentials of Bernal-stacked monolayer (SLG), bilayer (BLG), trilayer (TLG) and tetralayer graphene. The surface potentials monotonically increase with the increased thickness pf graphene layers, agreeing well with the previous work [43].

Once a twist angle is introduced into the bilayer graphene, the entire band structure and thus the Fermi level may change [44, 45]. Therefore, work functions are capable of serving as a fingerprint for characterizing tBLG domains with different twist angles. Fig. 2(f) and (g) show the simultaneously obtained AFM and KPFM images of one selected BLG region from the optical image. Two specific areas with different surface potentials are clearly resolved, but not in topography image, which is due to their different twist angles. The interior domain boundary of this bilayer graphene is highlighted by the while dashed lines. The typical wrinkles are also clearly resolved in Fig. 2(g), which may follow the polishing striations of the copper substrate and form during CVD-grown process. The cluster-like surface contaminations resulting from the transfer process were also resolved in Fig. 2(g). The cross-section topography and surface potential profiles are shown in Fig. 2(h). An approximate surface potential difference of ~27 mV was determined between the two BLG areas due to their different twist angles. Combining some previous work and Raman measurements, the AB-BLG and ABA-TLG regions are determined and named as BLG and TLG, respectively.

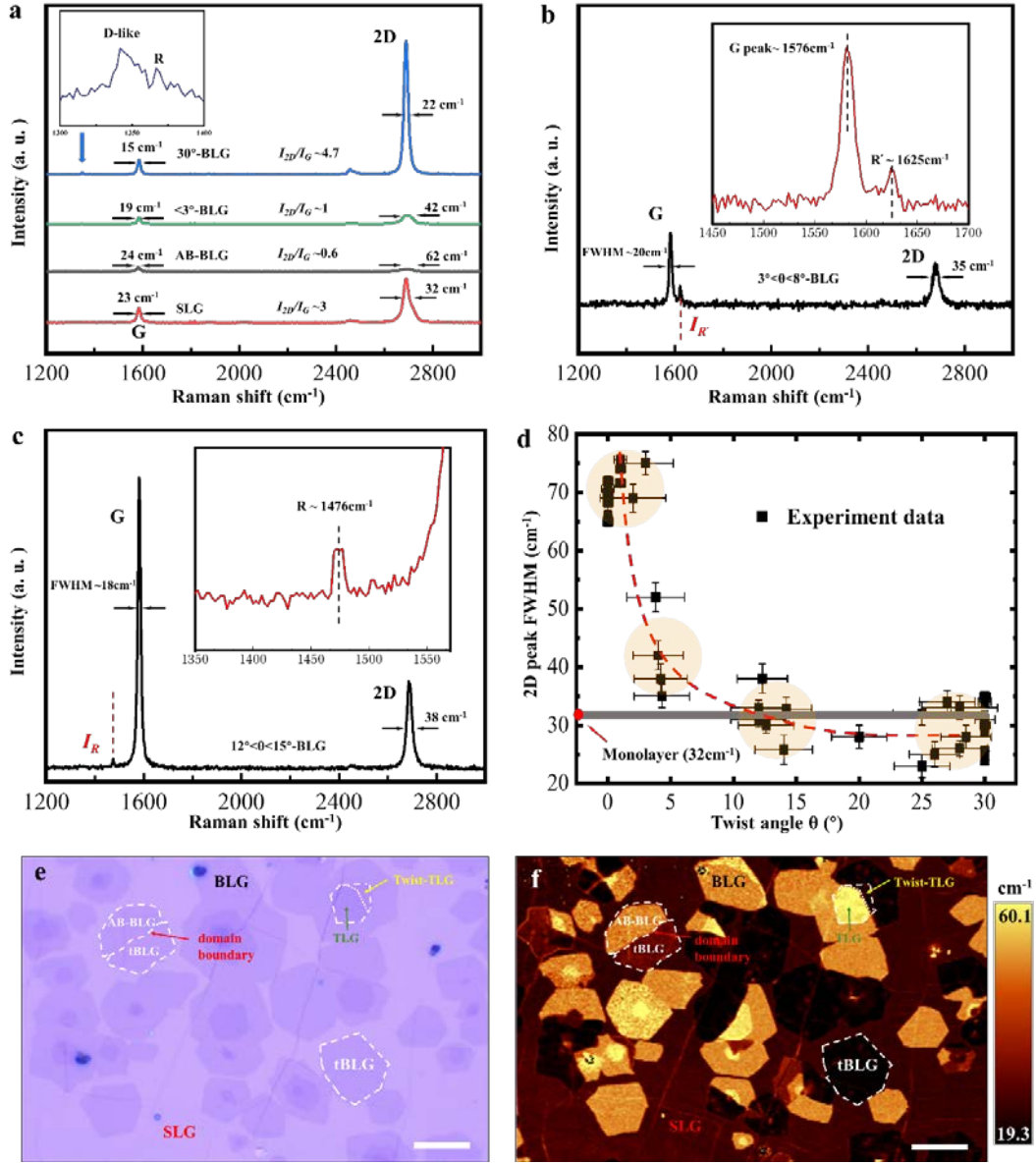

**Figure 3. Raman characterization of the CVD-grown bilayer graphene.** (**a**) Raman spectra of SLG and tBLG with different twist angles. Only wavenumber ranges near the G and 2D peaks are shown. The Raman spectrum of each curve has been vertically shifted for clarity. The red, black, green and blue curves represent the Raman spectra of SLG, AB-BLG, tBLG (twist angle, $\theta <3°$) and tBLG ($\theta \sim 30°$), respectively. The inset shows the zoom-in region of D-like peak and R peak in tBLG ($\theta \sim 30°$). (**b**) The Raman spectra of tBLG ($3°< \theta <8°$). Inset: the appeared Raman peak of R'-band at ~1625 cm$^{-1}$. (**c**) The Raman spectra of tBLG ($12°< \theta <15°$). Inset: the appeared Raman peak of R-band at ~1476 cm$^{-1}$. The $\theta$-dependent Raman spectral features of R- and R'-band result from the static interlayer potential-mediated inter- and intra-valley double-resonance Raman scattering processes. (**d**) Plot of 2D peak FWHM versus twist angle ($\theta$) in tBLG. The shaded circles of the experiment data collections are a guide to the eye. (**e**) Optical image of the CVD-grown multilayer graphene on the SiO$_2$/Si substrate. (**f**) The corresponding Raman 2D FWHM mapping acquired on the same area of (e). Scale bar is 30 µm in (e) and (f). The tBLG can exist as a single BLG flake or a part of BLG flake with interior tBLG and AB-BLG domains.

The Raman spectra measurements were further performed on these twisted multilayer graphene films, as shown in Fig. 3. Fig. 3(a) shows the Raman spectra of SLG, AB-BLG, tBLG ($\theta <3°$) and tBLG ($\theta \sim 30°$). The Raman modes for the G- and 2D-bands of SLG are located around 1576 cm$^{-1}$ and 2702 cm$^{-1}$, respectively. For the SLG, the intensity of G peak is much lower than that of 2D peak. While for the Bernal-stacked AB-BLG, the intensity of G peak is higher than that of 2D peak.

The parameter of 2D-to-G intensity ratio ($I_{2D}/I_G$) has been used to determine the twist angles of bilayer graphene [29-31]. For the CVD-grown graphene samples in this work, the twist angles of tBLG could be roughly grouped into four specific regimes of ($\theta <3°$, $3°< \theta <8°$, $12°< \theta <15°$, $\theta \sim 30°$). The 2D FWHM of tBLG ($\theta <3°$) is larger and smaller than that of SLG and AB-BLG, respectively. The origin of the larger 2D FWHM at this very-low-twist-angle regime is very complicated due to the local Bernal AB/BA stacking configurations resulting from the interfacial relaxation [30, 46]. The $I_{2D}/I_G$ ratio (and 2D FWHM) of the tBLG ($\theta \sim 30°$) is slightly larger (and smaller) than that of SLG, which have been attributed to their weak interlayer coupling. The emergence of D-like peak (~1348 cm$^{-1}$) and R peak (~1369 cm$^{-1}$) was attributed to their intrinsic quasicrystal states [47-49]. The emerged R'-band at ~1625 cm$^{-1}$ is unique for the tBLG ($3°< \theta <8°$), as shown in Fig. 3(b). For the tBLG ($12°< \theta <15°$), the characteristic R-band appears at the location of ~1476 cm$^{-1}$ in their Raman spectra, as shown in Fig. 3(c). In this twist angle regime, the plot the R-peak position as a function of the twist angle is also shown in Fig. S1, showing a similar trend with ref. [50]. The significant intensity of G-band in this regime is due to their enhanced Van Hove singularities (VHSs). Fig. 3(d) shows the obtained plot of 2D peak FWHM versus the twist angles in the four regimes of tBLG graphene. It is clearly that the tBLG prefer four specific twist angle regimes in our CVD-grown multilayer graphene films. Compare to the previous studies [51, 52], the sharp increasement of 2D FWHM at the twist angle of ($7°\sim9°$) was not observed due to the absence of these special tBLG in our work.

Fig. 3(e) and (f) show the corresponding optical image and Raman 2D FWHM map of the CVD-grown multilayer graphene films. The tBLG regions were clearly resolved in Fig. 3(f), which mainly exist as single BLG flake or a part of BLG flake with interior tBLG and AB-BLG domains. After a statistical analysis, we can conclude that the single flakes or domains of tBLG prefer the twist angle of ($\theta \sim 30°$) in our CVD-grown graphene films. The TLG flake with interior twist-TLG and ABA-stacked domains were also obtained, as shown in Fig.3(f). The detailed formation mechanism of these twisted graphene layers is still not very clear, which could be related with the underlying growth mode of multilayer graphene.

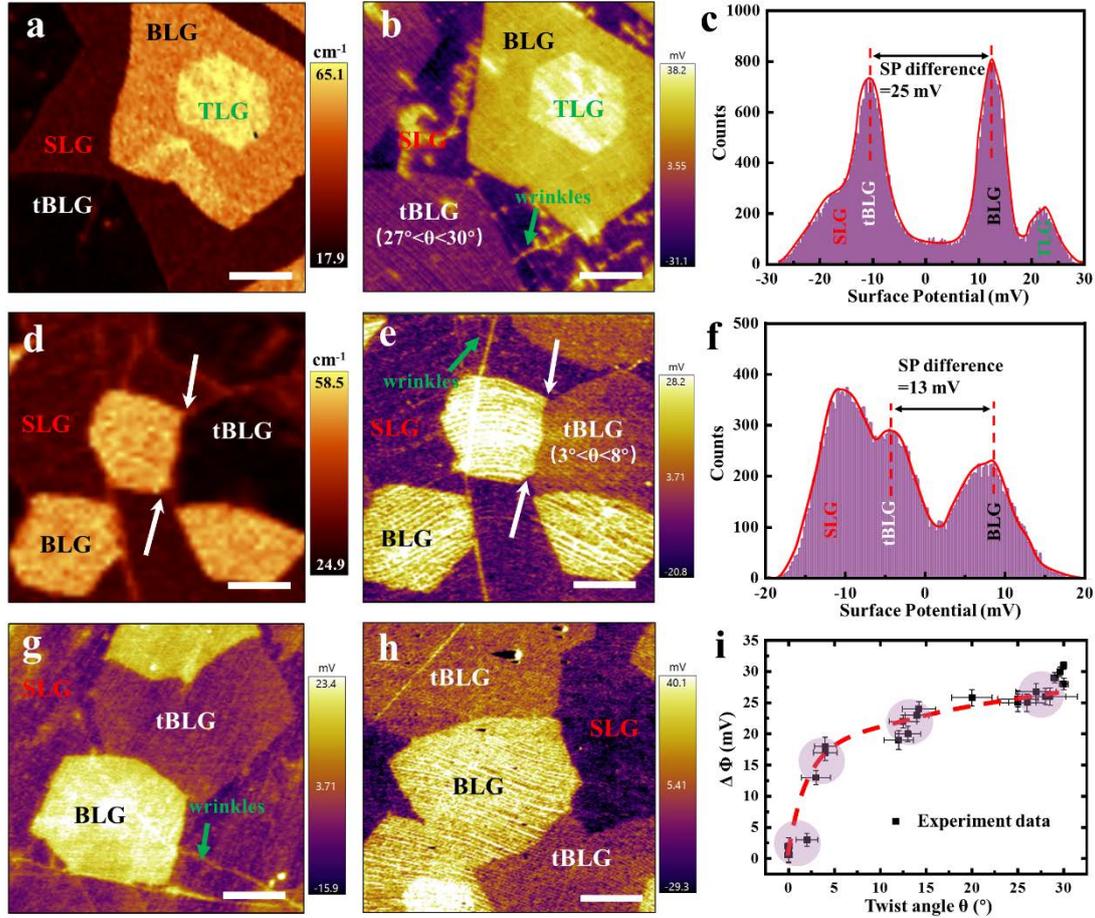

**Figure 4. Combined Raman and KPFM characterizations of the CVD-grown multilayer graphene.** (**a**) Raman 2D peak FWHM mapping of a conventional multilayer graphene region with a single tBLG flake. (**b**) The corresponding surface potential image of (a). (**c**) Histogram of surface potentials in (b), showing a potential difference of ~25 mV between tBLG and AB-BLG. (**d**) Raman 2D peak FWHM mapping of a bilayer graphene area with interior tBLG and BLG regions. (**e**) The corresponding surface potential image of (d). (**f**) Histogram of surface potentials in (e), showing a potential difference of ~13 mV between tBLG and AB-BLG. (**g, h**) Surface potential images of various bilayer graphene areas. (**i**) Plot of surface potential difference between tBLG and AB-BLG versus twist angle ($\theta$) in tBLG. The shaded circles of the experiment data collections are a guide to the eye. Scale bar is 6 μm in all images.

The *in-situ* combined Raman and KPFM measurements were further performed to correlate the various surface potentials and twist angles in tBLG, as shown in Fig. 4. Fig. 4(a) shows the Raman 2D peak FWHM mapping of Bernal-stacked multilayer graphene and a single tBLG region with a twist angle of (27°< $\theta$ <30°). The corresponding surface potential image shows the clear contrast between single tBLG and BLG domains, as shown in Fig. 4(b). The surface potential difference between tBLG (27°< $\theta$ <30°) and AB-BLG domains is determined at ~25 mV, as shown in the histogram of Fig. 4(c). Fig. 4(d) and (e) show the corresponding Raman 2D peak FWHM and KPFM surface potential mapping of one bilayer graphene region with interior tBLG (3°< $\theta$ <8°) and BLG domains. The surface potential difference between interior tBLG (3°< $\theta$ <8°) and AB-BLG domains is determined at ~13 mV, as shown in Fig. 4(f). In addition, the wrinkles of graphene layer were also clearly resolved in both Raman and KPFM mapping, and highlighted by green arrow. While the domain boundaries between tBLG and BLG (marked by white arrows) appear as a sharp change of contrast without any enhanced or decreased signal contrast in Raman and KPFM mapping. It can be concluded that the tBLG and BLG domain are atomically and smoothly merged without apparent mechanical strain [53-55].

More optical, SEM, Raman and KPFM measurement results of twisted bilayer graphene films were displayed in Fig. 2(g, h) and Fig. S(2-4). Fig. 4(i) shows the plot of surface potential (SP) difference $\Delta\Phi$ ($\Delta\Phi$=$SP_{BLG}$-$SP_{tBLG}$) between tBLG and AB-BLG versus twist angle($\theta$) in tBLG. Generally, it is clearly shown that the $\Delta\Phi$ of tBLG increased monotonically with the increased twist angle ($\theta$). For the tBLG ($\theta$ <3°), the $\Delta\Phi$ is pretty small, suggesting the strong energy-band coupling due to the tiny twist angles. Then the $\Delta\Phi$ steeply increased to ~15-20 mV ($\theta$~5°), and then slowly increased to ~25 mV at the twist angle of ($\theta$~30°), due to the gradually weaken interlayer coupling. For the tBLG ($\theta$~30°), the interlayer coupling is negligible. The tBLG ($\theta$~30°) could be as decoupled bilayer (SLG+SLG) graphene. It is also noted that the surface potential of tBLG ($\theta$~30°) is only slightly higher than that of SLG.

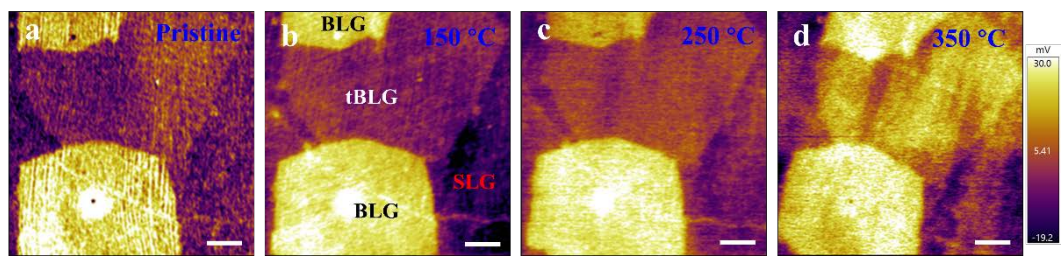

**Figure 5. Thermal annealing effect of twisted bilayer graphene (tBLG).** (**a-d**) Surface potential images of the typical tBLG regions taken by KPFM after a sequential of annealing processes in $H_2$/Ar atmosphere. The annealing temperatures were indicated in the corresponding images. The surface Scale bar: 6 µm.

The thermal annealing effect on the CVD-grown bilayer graphene films was further investigated by *in-situ* KPFM measurements after a sequential of annealing process in $H_2$/Ar atmosphere, as depicted in Fig. 5. The surface potential images of one typical tBLG region were sequentially displayed in Fig. 5(a-d), after annealing at 150°C, 250°C, 350°C for 2h and naturally cooling to the room temperature. It is noted that the graphene wrinkles were minimized due to the interfacial strain effect of the underlying $SiO_2$/Si substrate during the annealing process. The surface potential of AB-BLG show negligible change after annealing, which is in consistent with their most thermal stability. While the surface potential of tBLG is almost constant after the annealing at 250°C, but increased marginally after the following annealing at 350°C. It is indicated that the twist angle of tBLG could be apparently decreased via the interfacial sliding by the large thermal energy during the annealing process. It seems that the thermal stability of the tBLG is roughly dependent on the tBLG grain size [56, 57].

The ABA-TLG (Bernal-stacked), ABC-TLG (rhombohedral-stacked) and tTLG (twist-TLG) were also resolved in our KPFM and Raman measurements. Considering the emergent exotic quantum states within the semimetallic ABA-TLG and semiconducting ABC-TLG, the tTLG could be further investigated in the future work by KPFM. We also tried to obtain the Moiré pattern image of tBLG by other functional AFM methods, which could be due to the surface contaminations introduced during transfer process and the rough surface of $SiO_2$/Si substrate. The clean transfer method and the atomically flat h-BN will be needed to prepare the appropriate tBLG samples for visualizing their Moiré patterns. The in-plane homojunctions of BLG-tBLG could host novel transport properties, which is worthy of extensive exploring in the future.

In summary, by implementing KPFM measurements, we systematically studied the work functions dependence on the twist angles of the CVD-tBLG. We first showed that the surface potentials variations as a function of the thickness in Bernal-stacked graphene layers. KPFM images directly visualize the different stacking configurations of the AB-BLG and tBLG. The correlation of twist angles and surface potentials are further corroborated by KPFM mapping and Raman spectra. The results demonstrate that the surface potential difference $\Delta\Phi$ between AB-BLG and tBLG indicates an increased trend with an increasing twist angle (from 0° to 30°). In addition, we are able to visualize the surface potential variance of tBLG due to the effect of thermal annealing at high temperature. Our study thereby provides a comprehensive understanding as well as elucidates the twisted angle-dependent work functions of tBLG that could be helpful for exploring the evolution of twisted graphene systems.

**Acknowledgments:** This project is supported by the National Natural Science Foundation of China (NSFC) (No. 61674045), the Ministry of Science and Technology (MOST) of China (No. 2016YFA0200700), the Strategic Priority Research Program and Key Research Program of Frontier Sciences (Chinese Academy of Sciences, CAS) (No. XDB30000000, No. QYZDB-SSW-SYS031), Z. H. C. was supported by the Fundamental Research Funds for the Central Universities and the Research Funds of Renmin University of China (No. 21XNLG27).

## Materials and Methods

**Synthesis of CVD multilayer graphene on Cu foil.**

The multilayer graphene film was grown on copper foil, via low pressure chemical vapor deposition (CVD) method, as previous reports [15]. During the growth, 2 sccm methane ($CH_4$) and 30 sccm hydrogen ($H_2$) were introduced into the chamber at 1030°C for 40~90 minutes. After growth, the chamber was cooled down naturally to room temperature in $H_2$ (30 sccm). The as-grown graphene samples were transferred onto $SiO_2$/Si using a standard polymethyl-methacrylate (PMMA) transfer technique [35, 36]. Firstly, the graphene on Cu were spin-coated with a thin layer of PMMA. Next, the PMMA/graphene/Cu foil block was soaked in ammonium persulfate for 2h to etch the copper foil. Then the PMMA/graphene block was rinsed with deionized water and transferred onto a $SiO_2$/Si substrate. The residual PMMA was removed by acetone and isopropyl alcohol. Finally, the graphene/$SiO_2$/Si samples were then thermal annealed in an $H_2$/Ar atmosphere at 150°C for 2h.

**Raman and SEM measurements.**

The Witec 300R confocal Raman system was used to obtain the Raman spectroscopy and mappings. The excitation laser was tuned to wavelengths of 532 nm and 633 nm, the laser power was kept below 1~2 mW to avoid laser-induced sample heating or damage. The Witec Project 5.0 software was used for data analysis, creating a histogram of the integrated intensity, and achieving the Raman mapping. The morphology of BLG transferred on $SiO_2$/Si characterized by SEM with a Hitachi SU-5000 SEM with a voltage of 20 kV.

**AFM and KPFM measurements.**

The AFM and KPFM measurements were performed using an Asylum Cypher S AFM (Oxford Instruments-Asylum Research, Santa Barbara, USA). The AFM probe is a commercial electrostatic PPP-EFM (Nanosensors) with a resonance frequency of ~75 kHz. The KPFM measurement is performed via dual-pass mode. During the second pass scanning, the lift height is 20~30 nm. All the AFM measurements were conducted under ambient conditions (temperature, 23°C to 28°C; relative humidity, 20% to 30%).

# Twisted Angle-Dependent Work Functions in CVD-Grown Twisted Bilayer Graphene by Kelvin Probe Force Microscopy

Shangzhi Gu[1,2], Wenyu Liu[1], Guoyu Xian[2], Shuo Mi[1], Jiangfeng Guo[1], Songyang Li[1], Le Lei[1], Haoyu Dong[1], Rui Xu[1], Fei Pang[1], Shanshan Chen[1], Haitao Yang[2], Zhihai Cheng[1, *]

[1] *Beijing Key Laboratory of Optoelectronic Functional Materials & Micro-nano Devices, Department of Physics, Renmin University of China, Beijing 100872, China.*
[2] *Beijing National Laboratory for Condensed Matter Physics, Institute of Physics, Chinese Academy of Sciences, P.O. Box 603, Beijing 100190, China.*

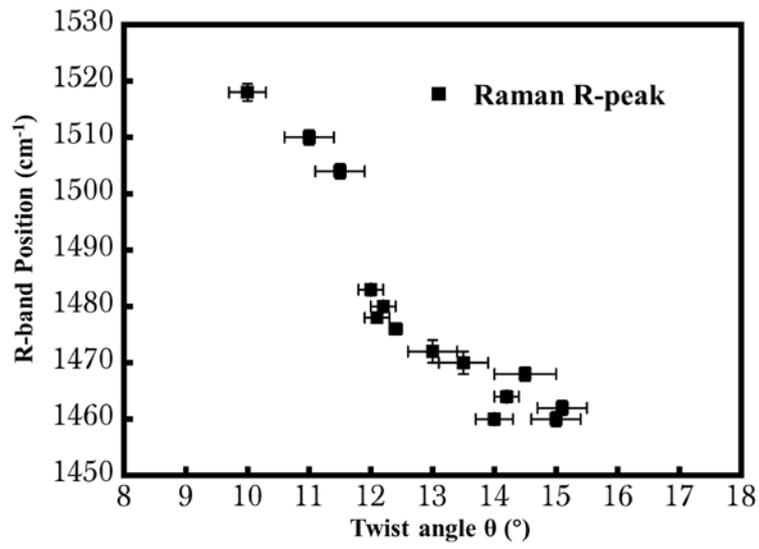

**Figure S1. The dependence of R-peak position on the twist angles for twisted bilayer graphene.** The R-peak position decreases with increasing twist angle from 10° to 15°.

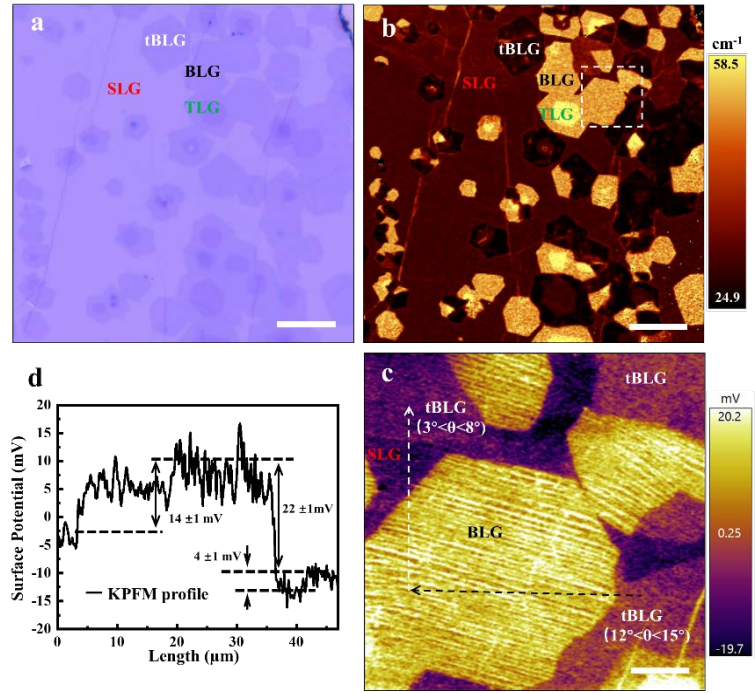

**Figure S2. Optical images and Raman maps of the CVD-grown twisted multi-layer graphene.** (**a**) Optical image of CVD-grown multilayer graphene transferred to the SiO$_2$/Si substrate, showing the continuing top graphene layer and the partial BLG (and TLG) in bottom layer. (**b**) The corresponding Raman 2D FWHM images acquired on the same area of (a). The SLG, BLG, tBLG (single or merging into BLG), TLG regions were clearly resolved by Raman 2D features. (**c**) Zoom-in KPFM image corresponding to (b) dashed white box. These bilayer graphene regions consist of the interior tBLG (3°< θ <8°) or tBLG (12°< θ <15°) and BLG domains. (**d**) The SP cross-section corresponding to the dashed black line in (c). Scale bar: (a-b) 30 µm, (c) 6 µm.

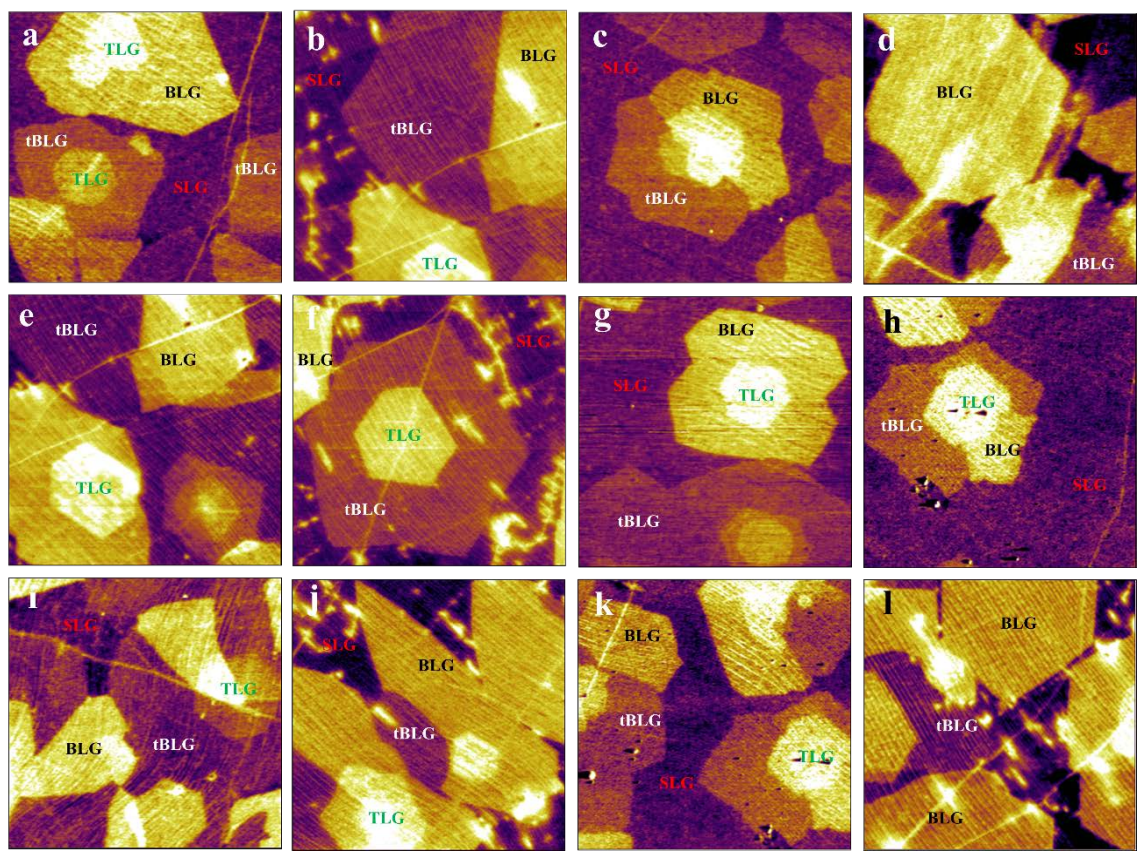

**Figure S3. Library of KPFM surface potential images of twisted graphene with various interlayer twist angles (0° ~ 30°) and domain structures.** (**a-l**) KPFM images of the multilayer graphene films. The surface potential domains of various irregular shape were observed. This result suggests that the domains may originate from the different mismatch angles. Images size: 30×30 µm.

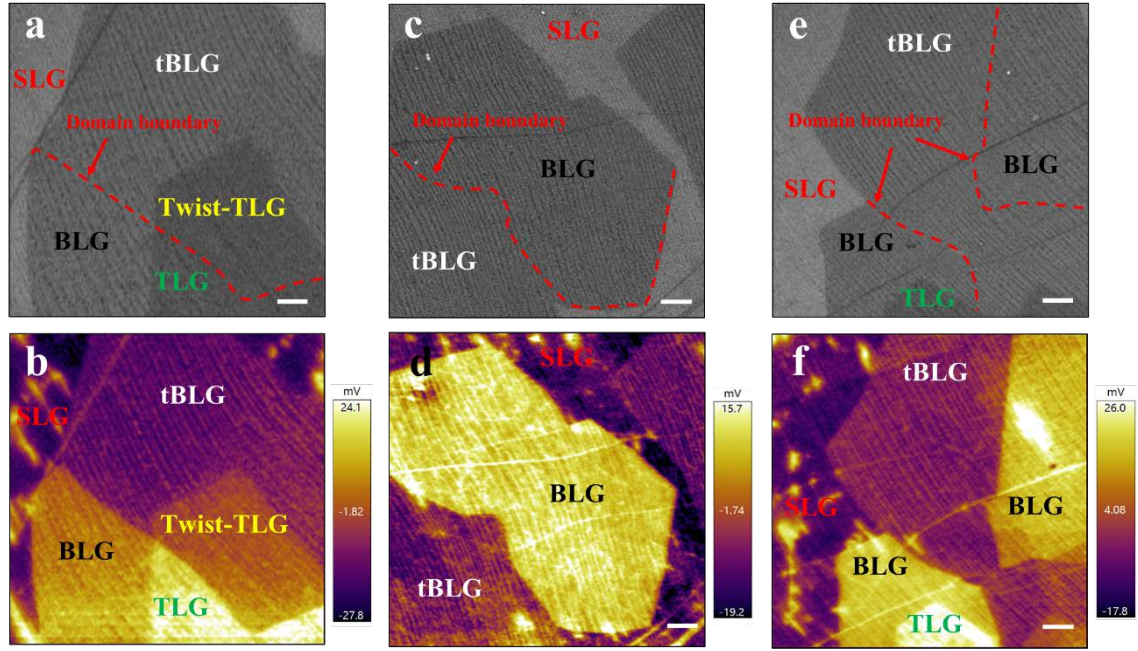

**Figure S4. SEM images and KPFM mapping of twisted bilayer graphene.** (**a, c, e**) SEM images of multi-layer uniform graphene films transferred on SiO$_2$/Si substrate. (**b, d, f**) The corresponding KPFM images SLG, BLG, tBLG, TLG, and twist-TLG in (a,c,e). The tBLG and BLG domains are clearly resolved based on their surface potential differences in the complete uniform BLG region in the SEM images, which are separated by the red dashed lines. Scale bar: 6 µm.